\documentclass[12pt]{article}

\usepackage{amsmath,amssymb}
\usepackage{eqnarray} 
\usepackage{cancel}
\usepackage{algorithm}
\usepackage{algpseudocode}
\usepackage{bm}
\usepackage{syllogism}

\usepackage[T1]{fontenc}
\usepackage{mathpazo,mathabx}
\usepackage[dvipsnames]{xcolor}

\usepackage{sectsty}
\sectionfont{\fontsize{12}{15}\selectfont}
\subsectionfont{\normalfont\fontsize{12}{15}\selectfont}
\subsubsectionfont{\itshape\normalfont\fontsize{12}{15}\selectfont}

\usepackage{lineno}

\usepackage[symbol]{footmisc}

\usepackage{relsize}
\newcommand{\E}{\mathop{\vcenter{\hbox{\relsize{+1}$E$}}}}

\DeclareMathOperator{\expit}{expit}

\usepackage{accents}
\newlength{\dhatheight}

\newcommand*\patchAmsMathEnvironmentForLineno[1]{%
	\expandafter\let\csname old#1\expandafter\endcsname\csname #1\endcsname
	\expandafter\let\csname oldend#1\expandafter\endcsname\csname end#1\endcsname
	\renewenvironment{#1}%
	{\linenomath\csname old#1\endcsname}%
	{\csname oldend#1\endcsname\endlinenomath}%
}
\newcommand*\patchBothAmsMathEnvironmentsForLineno[1]{%
	\patchAmsMathEnvironmentForLineno{#1}%
	\patchAmsMathEnvironmentForLineno{#1*}%
}
\AtBeginDocument{%
	\patchBothAmsMathEnvironmentsForLineno{equation}%
	\patchBothAmsMathEnvironmentsForLineno{align}%
	\patchBothAmsMathEnvironmentsForLineno{flalign}%
	\patchBothAmsMathEnvironmentsForLineno{alignat}%
	\patchBothAmsMathEnvironmentsForLineno{gather}%
	\patchBothAmsMathEnvironmentsForLineno{multline}%
}

\usepackage{color}
\usepackage{wrapfig}
\usepackage{indentfirst}
\usepackage{setspace}
\usepackage[numbers,super,sort&compress]{natbib}
\usepackage{booktabs}
\usepackage{rotating}
\usepackage[margin = .85in]{geometry}
\usepackage{fancyhdr}
\usepackage{tabularx}
\usepackage{pdfpages}
\usepackage{longtable}
\usepackage{multirow}
\usepackage{multicol}
\usepackage{float}
\usepackage{lscape}
\usepackage{graphicx}
\usepackage{mdwlist}
\usepackage{url}
\usepackage{comment}
\usepackage{amsthm}

\usepackage{caption}
\usepackage{subcaption}

\usepackage[compact]{titlesec}

\makeatletter
\renewcommand\@biblabel[1]{#1.}
\makeatother

\usepackage[colorlinks=true, urlcolor=blue, linkcolor=black, citecolor=black]{hyperref}

\graphicspath{{../figures/}}

\begin{document}
\thispagestyle{empty}
\baselineskip=28pt

\noindent

\vskip 0.2cm {{\noindent \huge Residual-on-Residual Regression as a Tool for Effect Estimation in Observational Data}}

\baselineskip=12pt

\vskip .25cm
\noindent Ashley I. Naimi, PhD$^{1,2 *}$\\[.5em]
\noindent Qianhui Jin, MPH$^{3}$\\[.5em]
\noindent Ya-Hui Yu, PhD$^{1}$\\[.5em]
\noindent Sara M. Parisi, MPH$^{3}$\\[.5em]
\noindent Lisa M. Bodnar, PhD$^{3}$\\[.5em]

\vskip .25cm
\noindent $^1$ Department of Epidemiology, Emory University.\\[.5em]
\noindent $^2$ Department of Data and Decision Science, Emory University.\\[.5em]
\noindent $^3$ Department of Epidemiology, University of Pittsburgh.\\[.5em]

\vskip .25cm
\noindent \hskip -.2cm
\begin{tabular}{ll}
*Correspondence: & Department of Epidemiology \\[-.1cm]
& Rollins School of Public Health \\[-.1cm]
& Emory University \\[-.1cm]
& 1518 Clifton Road \\[-.1cm]
& Atlanta, GA 30322\\[-.1cm]
& \href{mailto:ashley.naimi@emory.edu}{ashley.naimi@emory.edu}
\end{tabular}

\vskip .25cm
\noindent Conflicts: The authors have no conflicts to disclose.
\vskip .25cm
\noindent Acknowledgements: We thank Dr Brian Whitcomb for comments on a previous version of this manuscript.
\vskip .25cm
\noindent Funding: This study is supported by grant funding from the National Institutes of Health: R01 HD102313 and R01 HL174652 to Naimi AI and Bodnar LM. The study sponsor had no role in the study design; collection, analysis, and interpretation of data; writing the report; or the decision to submit the report for publication.

\vskip .25cm
\vfill
\noindent \hskip -.2cm
\begin{tabular}{ll}
Target Journal: 	   & \emph{American Journal of Epidemiology} (Practice of Epidemiology) \\[-.01cm]
Text word count: 	   & 3{,}959 / 4{,}000 \\[-.01cm]
Abstract word count:   & 197 \\[-.01cm]
Number of Figures: 	   & 3 \\[-.01cm]
Number of Tables:      & 1 \\[-.01cm]
Number of References:  & 46 \\[-.01cm]
Running head:  & Residual on Residual Regression  \\
\end{tabular}
%
%
\newpage
\thispagestyle{empty}
\begin{center}
{\Large{\bf Abstract}}
\end{center}
\baselineskip=12pt

\noindent
Epidemiologists increasingly use machine learning to adjust for high-dimensional confounding. Augmented inverse probability weighting (AIPW) and targeted maximum likelihood estimation (TMLE) are most widely used but may yield different results and both can become unstable under weak positivity violations. Residual-on-residual regression is a stable alternative that estimates an exposure effect encoded in a partially linear model by fitting confounder adjusted models for the outcome and exposure, then regressing outcome residuals against exposure residuals using ordinary least squares. We illustrate the approach using data from the Nulliparous Pregnancy Outcomes Study: Monitoring Mothers-to-Be (nuMoM2b; $n = 7{,}923$), estimating the association between high vegetable intake density and preeclampsia. Residual-on-residual regression, AIPW, and TMLE yielded concordant estimates, indicating a modest reduction in preeclampsia risk. In simulations, residual-on-residual regression was unbiased with near-nominal confidence interval coverage, performing comparably to AIPW and TMLE and substantially better than a misspecified parametric model when the exposure effect is approximately constant. However, in simulation settings with positivity violations, residual on residual regression outperformed AIPW and TMLE when the true effect was coded in a partially linear model. When the exposure effect is approximately constant, residual-on-residual regression is interpretable, computationally simple, and provides a triangulation strategy for observational causal inference.
%
%
%
\baselineskip=12pt
\par\vfill\noindent

\par\medskip\noindent
\newpage
\doublespacing
\setcounter{page}{1}

\section{Introduction}

Researchers routinely use observational data to estimate causal exposure effects.\cite{Naimi2023a} When standard parametric regression models are used, researchers must correctly specify the functional form of the confounder--outcome relationships and determine which interactions to include.\cite{Naimi2020a,Vansteelandt2022} In simple settings, where these might be known with some confidence, this is a viable approach. However, causal effect estimation is regularly carried out with high-dimensional data, where the number of potential confounders is large relative to the sample size.\cite{Mitra2022} Here, functional form and interaction decisions can be difficult to make correctly, and estimators based on misspecified models can introduce biases that standard sensitivity analyses may not detect.\cite{Naimi2023,Zivich2021}

Machine learning algorithms offer a data-adaptive alternative, flexibly modeling complex confounder-confounder and confounder-exposure interactions without requiring functional forms to be pre-specified.\cite{Hastie2009} Ensemble learners such as the Super Learner optimally combine predictions from a library of candidate algorithms.\cite{Laan2007} However, most machine learning methods optimize a loss-function with respect to a bias-variance tradeoff, introducing bias to offset the high variance that can arise under flexibility. If left unaccounted for, this bias can lead to poorly performing estimators of the causal effect for na\"{i}ve singly robust methods.\cite{Kennedy2024, Diaz2020}

Fortunately, optimal statistical properties and valid statistical inference can be recovered by deriving estimators within a semiparametric efficiency theory.\cite{Tsiatis2006} Such estimators are built around Neyman-orthogonal scores or influence functions of pathwise differentiable causal estimands, rendering them less sensitive to so-called ``first-order'' biases in nuisance estimation.\cite{Chen2026} The two approaches most widely used are augmented inverse probability weighting (AIPW)\cite{Robins1994} and targeted maximum likelihood estimation (TMLE).\cite{vanderLaan2006} Both are doubly robust in that they yield consistent estimators of the average treatment effect if at least one of the two nuisance functions (e.g., the outcome or propensity score model) is consistently estimated. When paired with some form of out-of-sample validation such as cross-fitting or cross-validation, both estimators can attain the nonparametric efficiency bound and support valid asymptotic inference.\cite{Kennedy2024}

However, these estimators are not a panacea for effect estimation. For instance, they can be particularly sensitive to even minor violations of the positivity assumption,\cite{Rudolph2022} which can occur when the sample size is not large enough to populate all confounder strata with exposed and unexposed individuals.\cite{Westreich2010,Petersen2012} This is a common problem in nutritional epidemiology, where researchers are often interested in the effects of ``healthy'' dietary interventions (such as consuming enough vegetables per 1,000 kcals of diet to meet federal guidelines) while adjusting for a complex confounder set. Few people tend to meet dietary guidelines for vegetable consumption,\cite{Bodnar2023a} leading to questionable positivity for the desired estimand. Additionally, even when positivity is met, applying both TMLE and AIPW to the same data can yield different results,\cite{Yadlowsky2022} leaving uncertainty about which estimate to rely on.

A less commonly used alternative to estimating causal effects with machine learning methods is a residual-on-residual approach, first introduced by Robinson in the late 1980s,\cite{Robinson1988} and more recently popularized in econometrics.\cite{Chernozhukov2018} Like AIPW and TMLE, it combines estimates of both outcome and exposure nuisance models. The approach is based on the specification of a partially linear model that encodes an average treatment effect under key assumptions.\cite{Robinson1988} It is particularly straightforward to implement and offers doubly-robust properties that complement those of AIPW and TMLE. Here, we illustrate residual-on-residual regression using an empirical and simulation example, provide R code to reproduce all results, and discuss key assumptions, strengths, and limitations of the residual-on-residual regression approach as a tool for estimating causal effects.

\section{Methods}

\subsection{Data Source}

We used data from the Nulliparous Pregnancy Outcomes Study: Monitoring Mothers-to-Be (nuMoM2b), a U.S.-based pregnancy cohort study conducted across eight clinical centers (2010 to 2013).\cite{Haas2016} A total of 10,038 nulliparous pregnant individuals (no prior pregnancy $\geq$20 weeks of gestation) with singleton pregnancies were enrolled at 6 to 13 weeks' gestation. At enrollment, trained research staff administered questionnaires on periconceptional diet, medical history, and demographic, psychosocial, behavioral, and environmental factors. Medical records were reviewed $\geq$ 30 days after delivery to obtain information on pregnancy and birth outcomes. All participants provided written informed consent, and institutional review boards at each site approved the protocol. The final analytic cohort included 7,923 participants after excluding 2,115 (21\%) with missing diet or outcome data.

\subsection{Exposure}

At enrollment, participants completed a self-administered food frequency questionnaire (FFQ), described previously,\cite{Bodnar2024} to assess usual dietary intake density (both frequency and portion size) during the three-month periconceptional period. NutritionQuest (Berkeley, CA) scanned FFQs and linked the line items to the Food Patterns Equivalents Database.\cite{USDAFoodNutrient2014} Food group units were divided by 1000 kcal of total energy to create intake densities.  We used data from the FFQ to construct our exposure, defined as high vegetable intake density ($\geq$ 1.25 cups/1000 kcal).

\subsection{Outcome}

We defined preeclampsia using the 2013 American College of Obstetricians and Gynecologists' diagnostic criteria,\cite{HypertensionPregnancyReport2013} adapted by the nuMoM2b investigators to the data collected.\cite{Facco2017}

\subsection{Covariates}

Confounders were all measured in the first trimester. They included study site, sociodemographics (e.g., age, education, race/ethnicity (proxy for structural racism), marital status, insurance), prepregnancy body mass index, preexisting diabetes or hypertension, smoking, physical activity,\cite{Piercy2018} psychosocial measures, percent poverty in the neighborhood, proximity to grocery stores, walkability,\cite{Giles-Corti2014} deprivation,\cite{Kind2018} a Healthy Eating Index (HEI) score, acculturation including maternal English language,  mother born in the US, gravidity, nausea and vomiting, whether the pregnancy was planned, binge drinking, and sleep factors. Additional details are provided in Appendix 1. Missing confounders were addressed using mean or mode imputation with indicators for missingness included in the models.

\subsection{AIPW and TMLE}

We first use AIPW and TMLE to estimate the average treatment effect, which can be encoded as a parameter in a nonparametric model:
\begin{equation}
Y = \mu(A, W) + \epsilon \label{eqn:nonpar}
\end{equation}
where $\mu(\bullet)$ is some nonparametric function, and where $E(\epsilon) = 0$. Throughout, we use $Y$ to denote the outcome, $A$ the exposure, and $W$ a confounding set. Under standard causal identification assumptions,\cite{Naimi2023a} this model implies:
$$\psi = E[Y^{a = 1} - Y^{a = 0}] = E[\mu(A = 1, W) - \mu(A = 0, W)]$$
where $Y^{a}$ is the outcome that would be observed if $A$ were set to value $a$. The right-hand side is the g-computation,\cite{Robins1986,Snowden2011} or marginal standardization,\cite{Naimi2020,Muller2014} plug-in estimator for a time-fixed exposure. Because $\psi$ is pathwise differentiable, it admits an efficient influence function whose empirical estimating equation, solved with nuisance functions fit to the data, yields a regular, asymptotically linear estimator attaining the semiparametric efficiency bound.\cite{Ross2024,Kennedy2024}

AIPW solves this estimating equation by augmenting the simple plug-in estimator with inverse-probability weights from an exposure model, a correction that removes first-order bias due to errors in either nuisance estimate.\cite{Kennedy2024} In contrast, TMLE is a substitution estimator that reaches the same solution indirectly, by tilting the initial outcome model predictions through a targeted fluctuation step parameterized by the propensity score model fit, until the estimating equation is satisfied, then plugging the updated predictions into the g-computation formula for $\psi$.\cite{Luque-Fernandez2018} Both AIPW and TMLE are well established methods extensively used in epidemiology.

\subsection{Residual-on-Residual Regression in a Partially Linear Model}

In contrast, residual-on-residual regression is less commonly deployed, though it shares some of the same doubly robust properties as AIPW and TMLE. We can define an effect of $A$ on $Y$ in the context of a \emph{partially} linear model such as:
\begin{equation}
Y = \psi A + g(W) + \epsilon \label{eqn:plm}
\end{equation}

In this equation, $g(W)$ is a nonparametric function denoting how the confounders relate to the mean of the outcome. This model is ``partially linear'' because it assumes that the exposure-outcome relationship is linear, but no such assumptions are made for the confounder-outcome relationships, which can be linear, polynomial, logarithmic, or any other smooth function. By making few assumptions about functional form, it lowers the chance of bias from model misspecification.

\subsection{The Intuition Behind Residual-on-Residual Regression}

Robinson first introduced a strategy to flexibly estimate $\psi$ encoded in this partially linear model.\cite{Robinson1988} His insight was based on the notion that, if we could remove the influence of the confounders $W$ from both $Y$ and $A$, we could then estimate $\psi$ by regressing the ``de-influenced'' $Y$ on the ``de-influenced'' $A$. The challenge is how to remove $W$'s influence without knowing the functional form of that influence. Robinson showed that it is possible to eliminate $g(W)$ altogether in equation \ref{eqn:plm} if we use outcome and exposure residuals in a simple linear model:
\begin{equation}
\big (Y - E[Y \mid W] \big )= \psi \big( A - E[A \mid W] \big) + \epsilon \label{eqn:ronr}
\end{equation}

\noindent where $E[Y \mid W]$ and $E[A \mid W]$ are models regressing $Y$ against $W$, and $A$ against $W$. At face value, this procedure seems to simply transfer the problem from equation \ref{eqn:plm} to the fits for these models. However, Robinson's main contribution was to show that we can estimate these models nonparametrically, for example, using flexible machine learning methods, and still obtain a residual-on-residual estimator of $\psi$ using standard OLS with optimal statistical properties.

The residual-on-residual regression approach proceeds in three steps. The first is to estimate a model regressing $Y$ on only confounders $W$ (no exposure). Using nonparametric regression, such as a stacking algorithm like the Super Learner, estimate the conditional mean of $Y$ given $W$. Denote the predictions from this model $\hat{m}_Y(W)$, and use them to compute the residuals: $\tilde{Y} = Y - \hat{m}_Y(W)$. Do the same for the exposure, fitting a regression model for $E[A \mid W]$, construct residuals $\tilde{A} = A - \hat{m}_A(W)$. Once these residuals are constructed, the next step is to estimate $\psi$ using a no-intercept ordinary least squares estimator regressing $\tilde{Y}$ against $\tilde{A}$. The resulting estimator is consistent and asymptotically normal, with optimal statistical properties despite using machine learning to estimate $\hat{m}_Y(W)$ and $\hat{m}_A(W)$.\cite{Robinson1988, Chernozhukov2018}

\subsection{Out-of-Sample Validation}

While residual-on-residual regression improves upon na\"{i}ve use of machine learning to estimate $\psi$, additional robustness can be gained using some form of out-of-sample validation,\cite{Zivich2022} including sample splitting, cross fitting, or cross validation. The simplest approach is to use cross-validation when fitting models for $m_Y(W)$ and $m_A(W)$. Importantly, the same individuals in each fold are used to train both algorithms. This creates a set of $K$ folds, each with predictions for $\hat{m}_Y(W)$ and $\hat{m}_A(W)$. Once created, these $K$ test folds can be combined back into a single dataset, residuals constructed, and the no-intercept OLS deployed.

\subsection{Variance Estimation}

Variance estimation can be straightforward for the residual-on-residual regression estimator. Robinson originally proposed the standard model-based (homoskedastic) OLS variance, easily obtained in standard software, which is valid under a conditional constant variance assumption.\cite{Robinson1988} Vansteelandt and Solander\cite{VansteelandtSjolander2016} and later Chernozhukov et al\cite{Chernozhukov2018} propose the robust (i.e., sandwich, or heteroskedasticity consistent) variance estimator, which can also be obtained from standard software. This is exactly the empirical variance of the estimator's Neyman-orthogonal influence function, on the same footing as the influence-function based standard errors used for AIPW and TMLE. We report the HC3 variant, which re-weights $\hat{\epsilon}_i^2$ by $(1-h_{ii})^{-2}$, where $h_{ii}$ is the hat matrix from the OLS fit, and is often the preferred member of the heteroskedasticity consistent (HC) family, particularly at modest sample sizes.\cite{Mansournia2021}

\section{Application to NuMoM2b Data}

We applied residual-on-residual regression, AIPW, and TMLE to estimate the association between high vegetable intake density ($\geq$ 1.25 cups per 1{,}000 kcal) and the risk of preeclampsia. Exposure and outcome models were estimated using a 10-fold cross-validated Super Learner ensemble, which included the sample mean, generalized linear models, random forests (ranger), regularized regression using the elastic net (glmnet), and multivariate adaptive regression splines (earth).

\begin{figure}[ht]
\centering
\includegraphics[width=0.75\textwidth]{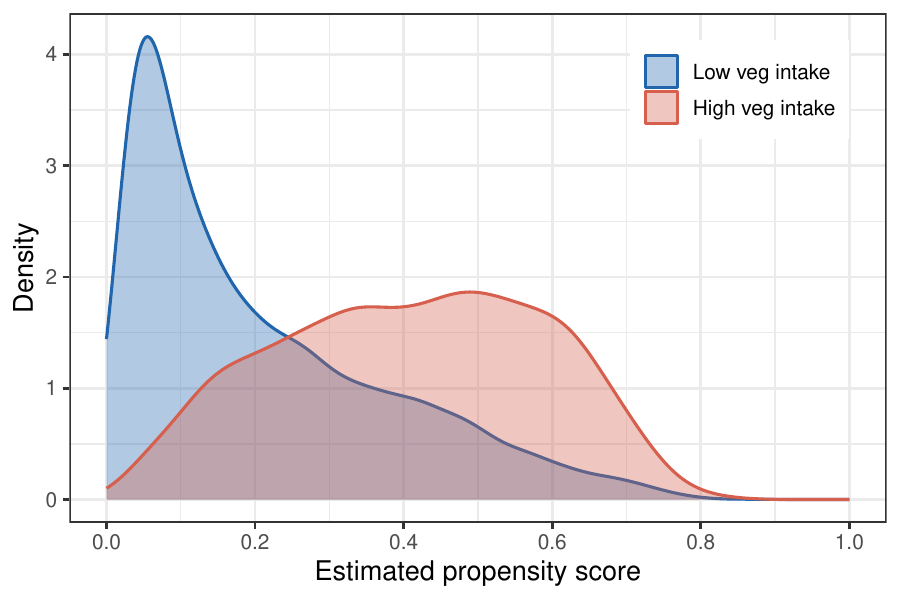}
\caption{Estimated propensity score distributions for the exposure (periconceptional high vegetable intake density, $\geq$ 1.25 cups per 1{,}000 kcal) in the Nulliparous Pregnancy Outcomes Study: Monitoring Mothers-to-Be (nuMoM2b; $n = 7{,}923$). Propensity scores were estimated using a 10-fold cross-fitted Super Learner ensemble of the sample mean, generalized linear models, random forests (ranger), elastic net (glmnet), and multivariate adaptive regression splines (earth). Densities are shown separately for participants with low ($<$ 1.25 cups per 1{,}000 kcal) and high vegetable intake density.}
\label{fig:ps_overlap}
\end{figure}

Among participants, 2{,}045 (25.8\%) had high vegetable intake density and 5{,}878 (74.2\%) had low vegetable intake density. A total of 682 (8.6\%) participants developed preeclampsia. The incidence of preeclampsia was higher among participants with low vegetable intake density (9.3\%) than those with high vegetable intake density (6.6\%). The distribution of estimated propensity scores by exposure status is shown in Figure~\ref{fig:ps_overlap}, and indicates fair overlap across the support of the propensity score, suggesting little to no violations of the positivity assumption. 

After adjustment for confounders, all three estimators indicated that high vegetable intake density was associated with fewer preeclampsia cases per 100 participants (Table~\ref{tab:numom_results}). Residual-on-residual regression yielded an adjusted risk difference (RD) of $-1.49$ per 100 (95\% confidence limits [CL]: $-2.94$, $-0.03$). AIPW and TMLE yielded comparable point estimates (AIPW RD: $-0.97$ per 100, 95\% CL: $-2.85$, $0.90$; TMLE RD: $-1.15$ per 100, 95\% CL: $-3.02$, $0.73$).

\section{Simulation Study}

We conducted a simple Monte Carlo simulation study to compare the finite-sample properties of residual-on-residual regression, AIPW, TMLE, and a misspecified linear-additive OLS model. For each simulated dataset we drew $n = 500$ independent and identically distributed observations on six baseline confounders $W = (W_1, \ldots, W_6)$, a binary exposure $A$, and a continuous outcome $Y$. The exposure followed a logistic propensity in $W$, and the outcome a partially linear model $Y = 125 + \psi A + g(W) + \varepsilon$ with $\varepsilon \sim \mathcal{N}(0, \sigma_Y^2)$; both the propensity and $g(W)$ contained quadratic and pairwise-interaction terms, and $W_6$ entered neither, serving as a noise covariate. Because $g(W)$ does not interact with $A$, the average treatment effect equals $\psi$ exactly. We swept $\psi \in \{-1, -0.5, 0, 0.5, 1\}$ crossed with $\sigma_Y \in \{0.5, 1, 2\}$, yielding 15 scenarios whose quadratic and interactive structure a linear-additive parametric model cannot recover. The full data-generating equations are given in Appendix~\ref{sec:sim_appendix}.

\subsection{Simulation Design}

We generated $10{,}000$ Monte Carlo replicates per scenario, with seeds shared across scenarios so that contrasts share Monte Carlo noise.\cite{Naimi2024b} Within each replicate, we estimated $\psi$ using residual-on-residual regression, AIPW, and TMLE, with nuisance functions fit by a 5-fold cross-validated Super Learner over a library of four generalized linear model variants: main effects only; main effects with all two-way interactions; main effects with quadratics for the continuous confounders; and a combined main-effects, quadratic, and interaction model. We opted to use these variants of a GLM instead of machine learning algorithms to lower the computational burden of our simulations. A naive linear-additive ordinary least squares (OLS) regression of $Y$ on $A$ and $W_1, \ldots, W_6$ was included in our simulations as a misspecified parametric comparator. Propensity score predictions were truncated to $[0.01, 0.99]$. Residual-on-residual regression was implemented via no-intercept OLS of outcome residuals on exposure residuals, with HC3 robust standard errors;\cite{Mansournia2021} AIPW and TMLE standard errors were obtained from the empirical standard deviation of the influence function,  evaluated on the cross-fitted nuisance estimates.

We summarized each estimator's performance across replicates using bias (the mean of $\hat{\psi} - \psi$), empirical standard error (Emp.~SE; the standard deviation of $\hat{\psi}$), average analytic standard error (Avg.~SE), root mean squared error (RMSE), and empirical 95\% Wald confidence interval coverage. We additionally report the standard error ratio Avg.~SE\,/\,Emp.~SE, for which values near unity indicate that the model-based standard error correctly captures the sampling variability of $\hat{\psi}$. The full replication code for the simulation study is available in the associated \href{https://github.com/ainaimi/ronr-illustration}{GitHub repository}.

\subsection{Positivity Simulations}

The primary simulations were conducted for scenarios where positivity is met. To evaluate each estimator's sensitivity to positivity violations, we ran a corresponding simulation that degrades overlap while holding the estimand fixed. We retained the same data-generating mechanism, but introduced a multiplier $\zeta \in \{1, 2, 3, 4, 6\}$ and replaced the coefficient of $-0.7$ for the relation between $W_2 \sim Uni(-1, 1)$ with $-0.7\zeta$ while holding all other terms in the propensity and outcome models fixed; $\zeta = 1$ reproduces the primary-simulation propensity, and larger $\zeta$ drives $\pi(W)$ toward $0$ and $1$, creating progressively more severe practical positivity violations. So that this instability would not be masked by aggressive clipping, we loosened the truncation of the estimated propensity to $[0.001, 0.999]$. All other settings were unchanged ($n = 500$, $\sigma_Y = 1$, $10{,}000$ replicates per scenario with common random numbers, and the same four-member Super Learner library). For each replicate we recorded the realized degree of positivity violation (proportion of the sample with true propensity in the tails, the smallest true propensity, and the largest inverse-probability weight).

\section{Simulation Results}

\subsection{Primary simulation}

\begin{figure}[ht]
\centering
\includegraphics[width=\textwidth]{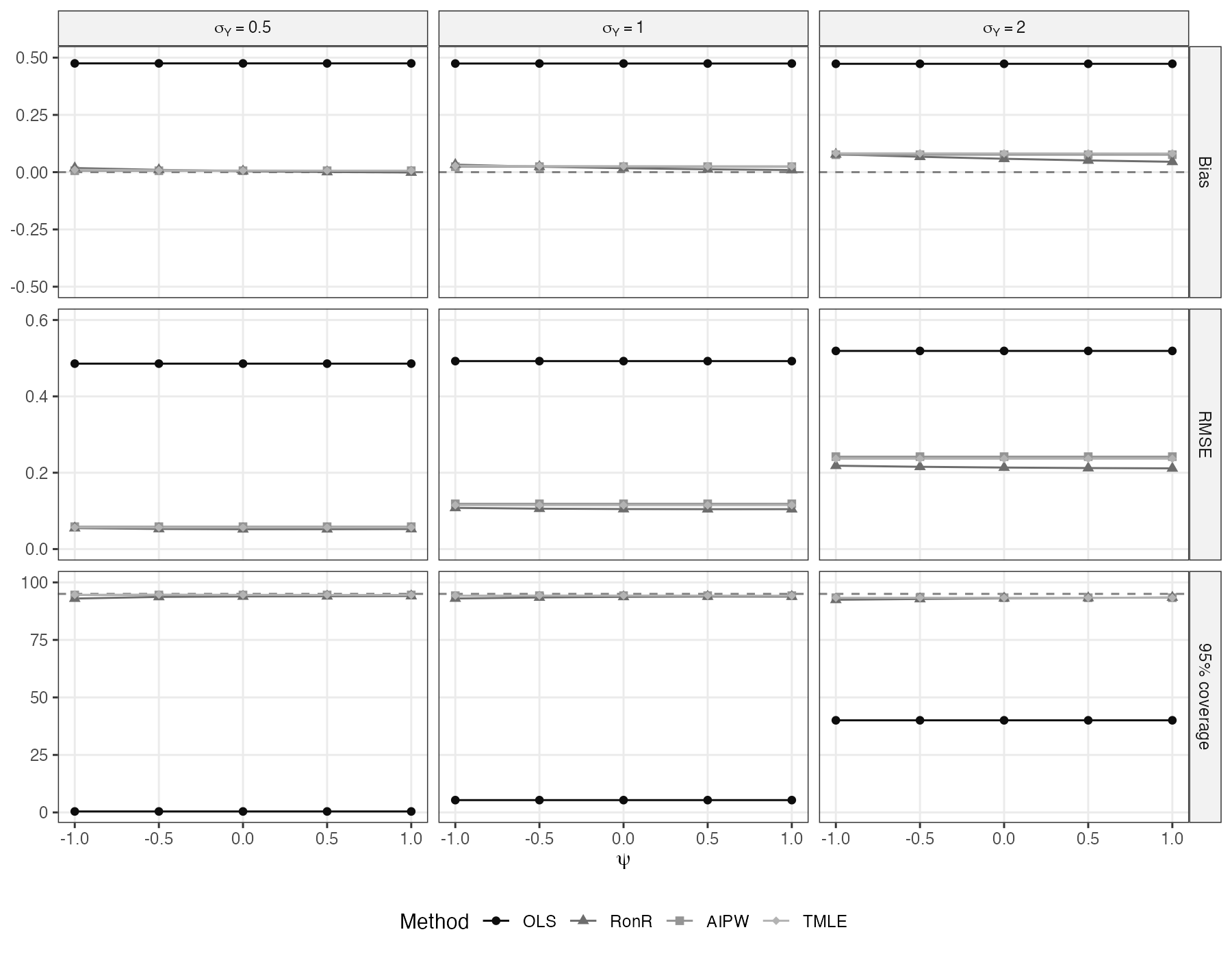}
\caption{Finite-sample performance of residual-on-residual regression (RonR), augmented inverse probability weighting (AIPW), targeted maximum likelihood estimation (TMLE), and a naive linear-additive ordinary least squares (OLS) comparator across $10{,}000$ Monte Carlo replicates per scenario at $n = 500$. Rows display bias (top), root mean squared error (RMSE; middle), and empirical 95\% Wald confidence interval coverage (bottom). Columns correspond to the outcome error scale $\sigma_Y \in \{0.5, 1, 2\}$. The horizontal axis shows the true treatment effect $\psi \in \{-1, -0.5, 0, 0.5, 1\}$. Dashed reference lines indicate bias $= 0$ and 95\% coverage. Nuisance functions for RonR, AIPW, and TMLE were fit by a 5-fold cross-validated Super Learner over a library of four generalized linear model variants spanning main effects, two-way interactions, and quadratic terms for the continuous confounders.}
\label{fig:sim_performance}
\end{figure}

Figure~\ref{fig:sim_performance} summarizes the finite-sample performance of residual-on-residual regression, AIPW, TMLE, and the naive linear-additive OLS comparator across the 15 simulation scenarios. As expected, the naive OLS estimator was biased in every scenario, and not dependent on the true treatment effect $\psi$ or the outcome error scale $\sigma_Y$ values. Because its linear-additive working model omits the quadratic and bilinear-interaction terms present in both the propensity and the outcome data-generating mechanisms, the coefficient for the treatment variable absorbed the corresponding omitted-variable bias. The model-based standard errors were correctly sized under the misspecified working model (SE ratio $\approx 1.00$), but (as expected) this structural bias overwhelmed the sampling variability of $\hat{\psi}$: 95\% Wald confidence interval coverage collapsed to $0.4\%$ at $\sigma_Y = 0.5$, $5.4\%$ at $\sigma_Y = 1$, and only $40.0\%$ at $\sigma_Y = 2$.

The three doubly robust estimators were, in contrast, essentially unbiased across the entire scenario grid. With $\sigma_Y \in \{0.5, 1\}$, the absolute bias of residual-on-residual regression never exceeded $0.04$, AIPW's bias was approximately $0.006$ at $\sigma_Y = 0.5$ and $0.024$ at $\sigma_Y = 1$, and TMLE's bias remained below $0.03$. At $\sigma_Y = 2$, bias rose modestly across all three doubly robust estimators to a maximum of approximately $0.08$, reflecting the increased difficulty of estimating $E[Y \mid W]$ and $E[Y \mid A, W]$ when the residual outcome scale is comparable to the magnitude of $g(W)$. Root mean squared errors of the doubly robust estimators were substantially smaller than that of the naive OLS comparator in every scenario, and the three doubly robust estimators were tightly clustered within each scenario (middle row of Figure~\ref{fig:sim_performance}).

95\% Wald confidence interval coverage for the doubly robust estimators was at or near the nominal level in every scenario: AIPW achieved $93.3$--$94.5\%$, TMLE achieved $93.3$--$94.7\%$, and residual-on-residual regression achieved $92.4$--$94.1\%$. The slight under-coverage of residual-on-residual regression corresponds to an SE ratio of approximately $0.96$--$0.98$, suggesting that the HC3 robust standard error mildly underestimates the empirical sampling variability of the no-intercept residual regression at $n = 500$; AIPW and TMLE produced SE ratios within $\pm 0.01$ of unity in every scenario. Because all three doubly robust estimators share the same cross-fitted Super Learner nuisance estimates, their estimates were correlated across simulation replicates; the small differences in bias and coverage visible in Figure~\ref{fig:sim_performance} reflect how each estimator combines these shared nuisance estimates.

\subsection{Positivity stress test}

\begin{figure}[ht]
\centering
\includegraphics[width=\textwidth]{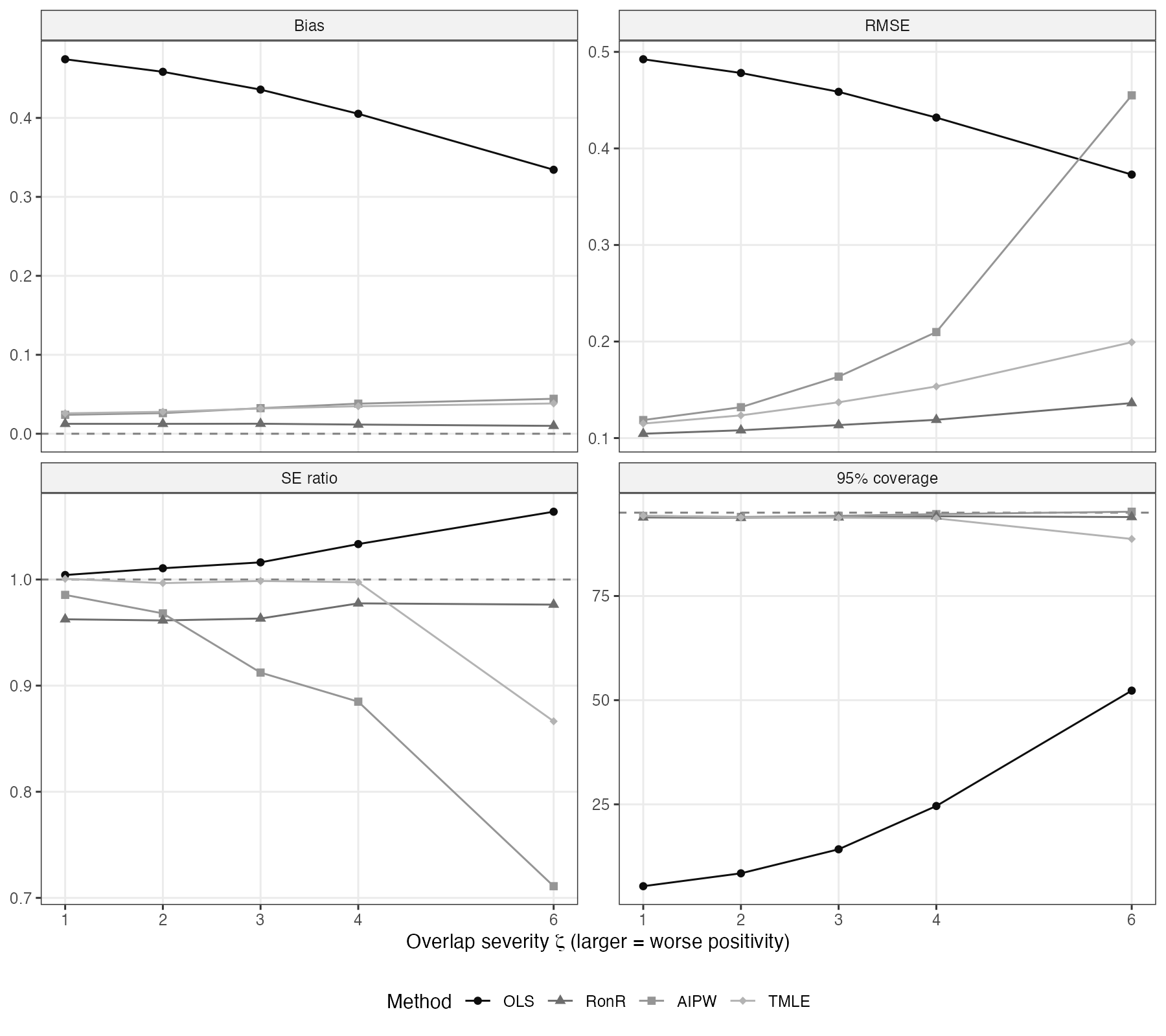}
\caption{Sensitivity to positivity violations under a homogeneous treatment effect ($\psi = 0.5$), across $10{,}000$ Monte Carlo replicates per scenario at $n = 500$. Because the effect is constant, the average treatment effect equals $\psi$ at every overlap level, so all estimators target the same parameter and any divergence reflects finite-sample instability. Panels show bias, root mean squared error (RMSE), the ratio of average analytic to empirical standard error (SE ratio), and empirical 95\% Wald confidence interval coverage. The horizontal axis is the overlap-severity dial $\zeta$; larger values steepen the propensity score's dependence on $W_2$, driving $\pi(W)$ toward $0$ and $1$, so that at $\zeta = 6$ roughly $30\%$ of the sample has true propensity outside $[0.05, 0.95]$. Dashed reference lines indicate bias $= 0$, SE ratio $= 1$, and $95\%$ coverage. RonR, residual-on-residual regression; AIPW, augmented inverse probability weighting; TMLE, targeted maximum likelihood estimation; OLS, naive linear-additive ordinary least squares.}
\label{fig:sim_positivity}
\end{figure}

As the overlap-severity dial $\zeta$ increased from $1$ to $6$, the realized degree of positivity violation grew substantially: the mean proportion of the sample with true propensity outside $[0.05, 0.95]$ rose from $1.8\%$ to $29.9\%$, the average smallest true propensity fell from $0.083$ to $0.008$, and the mean largest inverse-probability weight climbed from approximately $15$ to $105$. Against this backdrop, the estimators behaved differently (Figure~\ref{fig:sim_positivity}). Residual-on-residual regression was essentially invariant to positivity violatios: its bias stayed near $0.01$, its root mean squared error rose only modestly (from $0.10$ to $0.14$), and its coverage held between $93.8\%$ and $94.1\%$ throughout. AIPW, in contrast, incrementally lost precision: its root mean squared error roughly quadrupled (from $0.12$ to $0.46$) and its analytic standard error increasingly understated the empirical sampling variability, with the SE ratio falling from $0.99$ to $0.71$. Its Wald coverage nonetheless remained near nominal ($94$--$95\%$), because replicates with large estimation error also produced correspondingly large standard errors. TMLE performance was intermediate: the bounded fluctuation step guarded against the instability through $\zeta = 4$ (coverage $93.6\%$), but at the most severe setting its coverage fell to $88.7\%$ and its root mean squared error ($0.20$) exceeded that of residual-on-residual regression ($0.14$). The naive OLS comparator remained biased at every level for the same misspecification reason as in the primary simulation, independent of overlap.

\section{Discussion}

Modern semiparametric theory has produced several closely related strategies for estimating an average treatment effect from observational data using machine learning to flexibly adjust for confounders. Residual-on-residual regression takes an alternative route compared to standard AIPW or TMLE, assuming a linear structure for the exposure-outcome relationship, but leaving all other variables coded in the outcome model nonparametrically specified. This linear structure for the exposure has two implications: first, that there are no interactions between the exposure and other covariates in the model; second, for a continuous exposure, the dose-response relationship is linear. If the partially linear model with a constant treatment effect holds, all three estimators target the same parameter (i.e., are asymptotically equivalent) and, under cross-validation with appropriately fast nuisance convergence rates, are asymptotically linear with the same efficient influence function.\cite{Chen2026} 

Both the application and the simulation were consistent with this theoretical equivalence. In nuMoM2b, residual-on-residual regression, AIPW, and TMLE produced closely concordant estimates and supported the same substantive conclusion, with little indication of model dependence. The simulation extended the comparison to a setting with a known effect and deliberately nonlinear confounder relationships. There, the three doubly robust estimators behaved almost interchangeably, each remaining essentially unbiased with similar root mean squared error and coverage at or near the nominal level, while the deliberately misspecified linear-additive OLS comparator was substantially biased and severely under-covered. The one systematic difference among the doubly robust estimators was that the robust standard error for residual-on-residual regression was mildly anti-conservative at the sample size we studied, producing slightly below-nominal coverage, whereas the influence-function standard errors of AIPW and TMLE remained well calibrated. These findings illustrate in finite samples what asymptotic theory anticipates: residual-on-residual regression recovers the same effect as AIPW and TMLE at comparable efficiency, and all three flexible estimators markedly outperform a conventional parametric model that fixes the confounder functional form a priori.

Despite this asymptotic equivalence, residual-on-residual regression has practical advantages motivating its use alongside AIPW and TMLE. The closed-form coefficient and robust standard error make the estimator easy to implement in a wide variety of software platforms. Unlike AIPW, which forms explicit inverse-probability weights ($1/\hat\pi(W)$ and $1/\{1 - \hat\pi(W)\}$), and TMLE, whose targeting step uses a clever covariate that is an inverse function of the propensity score, residual-on-residual regression residualizes the exposure rather than inverting its estimated probability, which translates into greater stability when some propensity scores approach 0 or 1. Vansteelandt and Sjolander emphasize this point in the context of the closely related g-estimation approach, noting that it requires only a model for the conditional expectation of the exposure (as opposed to its conditional density).\cite{VansteelandtSjolander2016} Because of its close relationship to g-estimation, the simplicity and stability of residual-on-residual regression makes it a natural triangulation tool: any disagreement between AIPW, TMLE, or residual-on-residual estimates points to potential model dependence that warrants further investigation.

The interpretability of residual-on-residual regression as an estimator of the average treatment effect, however, hinges on the assumption that the treatment effect is constant across the population. When this assumption is violated, the estimator still converges to a meaningful causal quantity: Lal and Chou show that with a binary exposure and heterogeneous unit-level effects $\theta_i$, the probability limit is the conditional variance-weighted average $\E[\omega_i \theta_i]$, with weights $\omega_i = (A_i - \pi(W_i))^2 / \E[(A_i - \pi(W_i))^2]$, so that units with propensity scores closest to $0.5$ contribute most strongly and those nearest the boundaries least.\cite{LalChou2026} Like g-estimation, then, residual-on-residual regression is relatively insensitive to positivity violations. Our stress test makes this concrete: as overlap degraded to the point that nearly a third of the sample had a true propensity outside $[0.05, 0.95]$, residual-on-residual regression was essentially unchanged in bias, precision, and coverage, whereas AIPW's root mean squared error quadrupled and TMLE began to under-cover at the most severe setting.

This same weighting, however, makes the estimand depart from the average treatment effect under heterogeneity, by exactly $\text{Cov}(\omega_i, \theta_i)$, which vanishes only when effects are homogeneous or the propensity is constant.\cite{LalChou2026} Residual-on-residual regression is, in this light, a g-estimator of a constant-effect structural mean model, the simplest member of a broader family;\cite{VansteelandtSjolander2016} any effect modification must be encoded explicitly in that model to be recovered without bias, whereas AIPW and TMLE target the average treatment effect directly through the efficient influence function and do not share this bias. Appendix~\ref{sec:ronr_gest} gives the explicit large-sample limits, including the variance-weighted structural-mean-model form and the continuous-treatment case.

Residual-on-residual regression also forms the foundation for several recent extensions to heterogeneous treatment effects. Notably, the R-learner generalizes the ``partialling-out'' idea to estimate conditional average treatment effects such as $\tau(W) = \E[Y(1) - Y(0) \mid W]$ by minimizing a regularized empirical version of a loss function very similar to the residual-on-residual regression approach, but over a flexible function class for $\tau$.\cite{Nie2021} Thus, while we demonstrate its use in the context of estimating average treatment effects encoded in a partially linear model, the approach can be adapted to evaluate effect modification.

Our simulations were designed for illustration, not to deeply evaluate the estimators across a range of settings, and have several limitations. We relied on a relatively simple, narrow set of data generating mechanisms, with simple machine learning methods (a stacked generalization of four generalized linear models) used to fit the nuisance functions. By construction, the treatment effect in all our simulations is homogeneous, so we did not evaluate the expected bias of the residual-on-residual approach when the true ATE is heterogeneous. Indeed, the same down-weighting of near-deterministic units that stabilizes residual-on-residual regression is what makes its estimand depart from the true ATE when effects are heterogeneous, leading to potential bias. Additionally, positivity violations in our simulations were induced through a single mechanism (multiplying the role that a continuous confounder played in the propensity score model) at a single sample size, and with a single outcome model, so the precise magnitude of the AIPW and TMLE degradation is specific to this design. The instability is also sensitive to the truncation threshold: more aggressive clipping would temper the AIPW and TMLE variance at the cost of additional bias. 

In the partially linear model with an approximately constant treatment effect, residual-on-residual regression delivers an interpretable, computationally efficient, and stable estimator that complements AIPW and TMLE, and provides a natural triangulation strategy for observational causal inference. When the constant treatment effect assumption is suspect, residual-on-residual regression should be paired with an explicit assessment of effect heterogeneity, possibly through stratified analyses or estimation of conditional average treatment effects via the R- or other ``learners''.\cite{Kennedy2023} Though underutilized, it can benefit researchers estimating effects with machine learning.

\newpage

\begin{table}[ht]
\centering
\caption{Adjusted risk difference (per 100 participants) for the association between periconceptional high vegetable intake density ($\geq$ 1.25 cups per 1{,}000 kcal) and preeclampsia, estimated by residual-on-residual regression, augmented inverse probability weighting (AIPW), and targeted maximum likelihood estimation (TMLE) in the Nulliparous Pregnancy Outcomes Study: Monitoring Mothers-to-Be (nuMoM2b; $n = 7{,}923$). All estimators were fit with a 10-fold cross-fitted Super Learner ensemble, adjusting for sociodemographic, behavioral, medical, neighborhood, and dietary confounders. CL, confidence limit.} 
\label{tab:numom_results}
\begin{tabular}{lrrr}
  \toprule
Method & Risk Difference (per 100) & Lower 95\% CL & Upper 95\% CL \\ 
  \midrule
Residual-on-Residual & -1.49 & -2.94 & -0.03 \\ 
  AIPW & -0.97 & -2.85 & 0.90 \\ 
  TMLE & -1.15 & -3.02 & 0.73 \\ 
   \bottomrule
\end{tabular}
\end{table}

\newpage

\bibliographystyle{epid}
\bibliography{main_references}

\newpage

\appendix

\section{Confounder Selection}

Participants self-reported sociodemographic characteristics including age, education, race / ethnicity, marital status, public insurance status, nativity, English language proficiency, acculturation, and partner's education. Racial / ethnic identity had four mutually exclusive groups and was used as a proxy for exposure to structural racism.\cite{Ukoha2022} Prepregnancy BMI was calculated from self-reported prepregnancy weight and measured height at enrollment. Gravidity and whether the pregnancy was planned were self-reported. Pre-existing medical conditions including chronic hypertension, diabetes, sleep apnea risk, and thyroid medication use were ascertained from medical records. Standardized assessment tools\cite{Haas2015} measured depressive symptoms (Edinburgh Postnatal Depression Scale), perceived stress, anxiety, average nightly sleep duration, insomnia severity, and sleep satisfaction during the first trimester; nausea and vomiting severity was assessed using the Pregnancy Unique Quantification of Emesis (PUQE) score. Prepregnancy smoking status, alcohol use during pregnancy, and weekly alcohol consumption in the three months before pregnancy were self-reported. Physical activity in the 4 weeks before enrollment was assessed as total metabolic-equivalent task (MET) minutes per week.\cite{Piercy2018} Neighborhood characteristics such as the percentage of residents living below the federal poverty line, distance to nearby grocery stores, neighborhood walkability,\cite{Walker2013} and the area deprivation index,\cite{Kind2018} were derived from U.S. Census data linked to each participant's residential census tract and block group at enrollment. To account for the multidimensional nature of dietary patterns, overall diet quality was captured by the Healthy Eating Index 2010 (HEI-2010) total score,\cite{Shams-White2023} computed from the same food frequency questionnaire used to measure vegetable intake density. Study site was included to account for potential between-center differences.

\section{Simulation Study: Complete Implementation Details}\label{sec:sim_appendix}

This appendix provides a complete specification of the Monte Carlo simulation study reported in the Simulation Study and Simulation Results sections of the main paper. The full R replication code is in \texttt{code/simulation\_study\_sl.R} of the project's associated \href{https://github.com/ainaimi/ronr-illustration}{GitHub repository}, which also contains the seed bank used and the script that produces Figure~\ref{fig:sim_performance}.

\subsection*{A.1\quad Data-generating mechanism}

For each Monte Carlo replicate we drew an independent and identically distributed sample of size $n = 500$ from the following joint distribution. The confounders $W_1$ and $W_2$ were independently drawn from a Uniform$(-1, 1)$ distribution, $W_3$ and $W_4$ from a standard Normal distribution, and $W_5$ and $W_6$ from independent Bernoulli distributions with success probabilities $0.4$ and $0.5$, respectively. The variable $W_6$ entered neither the propensity nor the outcome model and serves as a noise covariate. The binary exposure was drawn from $A_i \sim \mathrm{Bernoulli}\{\pi(W_i)\}$, where $\pi(W) = \expit\{\eta(W)\}$ and
$$\eta(W) = -0.3 + 0.9 W_1 + 1.0 W_1^2 - 0.7 W_2 + 0.8 W_2 W_3 - 0.5 W_3 + 0.6 W_5 - 0.4 W_5 W_4.$$
The continuous outcome was generated from a partially linear model:
$$Y_i = 125 + \psi A_i + g(W_i) + \varepsilon_i, \qquad \varepsilon_i \sim \mathcal{N}(0, \sigma_Y^2),$$
where
$$g(W) = 1.2 W_1 + 1.5 W_1^2 - 1.0 W_2 + 1.4 W_2 W_3 - 0.8 W_3 + 0.7 W_4 + 1.0 W_5 - 0.6 W_5 W_4.$$
Because $g$ does not interact with $A$, the population average treatment effect equals $\psi$ exactly. We swept $\psi \in \{-1, -0.5, 0, 0.5, 1\}$ crossed with $\sigma_Y \in \{0.5, 1, 2\}$, yielding 15 scenarios. The intercept of $125$ is cosmetic and affects neither bias nor coverage; it places the outcome on a scale resembling the application data.

\subsection*{A.2\quad Monte Carlo replication and common random numbers}

We performed $10{,}000$ Monte Carlo replicates per scenario. Seeds were taken from a fixed reproducible seed bank (\texttt{data/random\_seed\_values.csv}) and held fixed across scenarios so that contrasts between scenarios share Monte Carlo noise (common random numbers). The Monte Carlo standard error on the reported bias is therefore approximately the empirical standard error divided by $\sqrt{10{,}000} = 100$, which is small relative to the entries shown in Figure~\ref{fig:sim_performance}.

\subsection*{A.3\quad Nuisance estimation}

Within each replicate we estimated three nuisance functions: $\pi(W) = P(A = 1 \mid W)$, $\mu(A, W) = \E[Y \mid A, W]$, and $\nu(W) = \E[Y \mid W]$. Each was fit by \texttt{CV.SuperLearner} from the R package \texttt{SuperLearner} using a 5-fold outer cross-validation. The same outer-fold partition was shared across the three nuisance fits by passing identical \texttt{validRows} indices, ensuring cross-fitted predictions for all three nuisances are aligned by observation index. Within each outer training partition, ensemble weights were chosen by an inner 5-fold cross-validation using non-negative least squares (\texttt{SuperLearner}'s default for Gaussian outcomes; logistic for binary).

The Super Learner library comprised four generalized linear model variants:
\begin{enumerate}
\item \texttt{SL.glm}: main effects in the input columns only.
\item \texttt{SL.glm.interaction}: main effects plus all two-way interactions between input columns.
\item \texttt{SL.glm.quad}: main effects plus a quadratic term for each continuous column (identified at fit time as a numeric column with more than two unique values not all contained in $\{0,1\}$).
\item \texttt{SL.glm.full}: main effects, quadratic terms for the continuous columns, and all two-way interactions between the original input columns.
\end{enumerate}
Only \texttt{SL.glm.full} is correctly specified for the data-generating mechanism above. Including the simpler members forces the Super Learner to discover this empirically through inner cross-validation rather than by oracle assignment. For $\mu(A, W)$, the design matrix passed to each library member included $A$ as the first column; the interaction- and full-models therefore also estimate $A \cdot W$ interactions, which are zero in expectation by construction.

Counterfactual outcome predictions $\hat\mu_0(W_i)$ and $\hat\mu_1(W_i)$ were obtained by, for each outer fold $k$, predicting on the validation rows of fold $k$ with $A$ forced to $0$ and $1$ in turn, using the Super Learner ensemble trained on the remaining four folds. Predictions were stored by positional assignment so that observation ordering is preserved. The propensity-score predictions were truncated to $[0.01, 0.99]$ prior to use in any estimator.

\subsection*{A.4\quad Estimator implementations}

The four estimators reported in Figure~\ref{fig:sim_performance} were implemented as follows.
\begin{itemize}
\item \emph{Naive OLS comparator.} Fit by \texttt{lm(Y \textasciitilde~A + C1 + \ldots + C6)} on the simulated data, with the model-based standard error from \texttt{summary(fit)\$coefficients}. This estimator is deliberately misspecified relative to the data-generating mechanism (it omits $W_1^2$, $W_2 W_3$, and $W_5 W_4$).
\item \emph{Residual-on-residual regression.} Fit by no-intercept ordinary least squares of $\tilde Y = Y - \hat\nu(W)$ on $\tilde A = A - \hat\pi(W)$, with HC3 robust standard errors from the \texttt{sandwich} package.
\item \emph{AIPW.} Computed as the sample mean of the efficient influence function
$$\varphi(O; \hat\pi, \hat\mu) = \hat\mu_1(W) - \hat\mu_0(W) + \frac{A \{Y - \hat\mu_1(W)\}}{\hat\pi(W)} - \frac{(1 - A)\{Y - \hat\mu_0(W)\}}{1 - \hat\pi(W)},$$
with standard error equal to the sample standard deviation of $\varphi$ divided by $\sqrt{n}$.
\item \emph{TMLE.} Computed by the \texttt{tmle} R package with $\mathrm{family} = \texttt{"gaussian"}$, the cross-fitted initial outcome predictions supplied via $Q = (\hat\mu_0, \hat\mu_1)$, the cross-fitted truncated propensity supplied via $g_{1W} = \hat\pi$, and the standard error taken from the package's targeted influence-function variance.
\end{itemize}

\subsection*{A.5\quad Software and reproducibility}

All computation was performed in R using the \texttt{SuperLearner}, \texttt{tmle}, \texttt{sandwich}, \texttt{lmtest}, \texttt{future.apply}, and \texttt{progressr} packages. Replicates were parallelized across CPU cores using a \texttt{multisession} \texttt{future} plan with per-replicate seeding managed by \texttt{future.seed = TRUE}; this keeps draws within a replicate reproducible while allowing the work to be distributed. Per-replicate Super Learner ensemble weights for each of $\pi$, $\mu$, and $\nu$ were recorded alongside the point estimates and standard errors, supporting post hoc diagnosis of which library members the ensemble was relying on for each scenario. The complete simulation pipeline is contained in \texttt{code/simulation\_study\_sl.R}; the figure-producing block at the end of that script writes \texttt{output/figures/sim\_performance\_sl.png}, which is reproduced as Figure~\ref{fig:sim_performance} of the main paper.

\subsection*{A.6\quad Positivity stress test}

The positivity stress test reported in the main text uses the same machinery as the primary simulation, with three changes, and is implemented in \texttt{code/simulation\_study\_sl\_positivity.R}. First, the treatment effect is homogeneous ($\psi = 0.5$ for every unit, with no exposure-by-confounder interaction), so the average treatment effect equals $\psi$ at every level of overlap. Second, the propensity score's dependence on $W_2$ is steepened by a factor $\zeta \in \{1, 2, 3, 4, 6\}$: its coefficient becomes $-0.7\zeta$, with all other terms in $\eta(W)$ and the entire outcome model held fixed, so that $\zeta = 1$ reproduces the primary-simulation propensity. Third, estimated propensity scores are truncated to $[0.001, 0.999]$ rather than $[0.01, 0.99]$. All other settings ($n = 500$, $\sigma_Y = 1$, $10{,}000$ replicates per scenario, common random numbers, and the four-member Super Learner library) are unchanged. Each replicate additionally records the realized degree of positivity violation (the tail mass and extremes of the true propensity and the largest inverse-probability weight) and a TMLE convergence flag. The figure-producing block writes \texttt{output/figures/sim\_performance\_sl\_positivity.png}, reproduced as Figure~\ref{fig:sim_positivity} of the main paper, together with summary tables of performance, realized positivity violation, and Super Learner ensemble weights by $\zeta$.

\section{Large-Sample Behavior of Residual-on-Residual Regression Under Effect Heterogeneity}\label{sec:ronr_gest}

When the constant-effect assumption underlying the partially linear model in equation~\ref{eqn:plm} fails, residual-on-residual regression remains interpretable as a g-estimator of a structural mean model, but its probability limit departs from the average treatment effect in ways that have been characterized precisely.

For a binary exposure with heterogeneous unit-level effects $\theta_i = Y_i(1) - Y_i(0)$, Lal and Chou\cite{LalChou2026} show that the probability limit of residual-on-residual regression is the conditional variance-weighted average $\E[\omega_i \theta_i]$, with weights $\omega_i = (A_i - \pi(W_i))^2 / \E[(A_i - \pi(W_i))^2]$. Relative to the true average treatment effect $\E[\theta_i]$, the bias is exactly $\text{Cov}(\omega_i, \theta_i)$, which vanishes only when effects are homogeneous or the treatment is assigned with constant probability across the population. Vansteelandt and Sjolander\cite{VansteelandtSjolander2016} reach the same conclusion within a structural mean model framework: when the constant-effect assumption is misspecified but the propensity is correctly modeled, the linear-SMM g-estimator that is asymptotically equivalent to residual-on-residual regression converges to $\E[\text{Var}(A \mid W) \, \{Y(1) - Y(0) \mid W\}] / \E[\text{Var}(A \mid W)]$, a variance-weighted average of conditional effects. For continuous treatments with a nonlinear dose-response, Lal and Chou\cite{LalChou2026} further show that the bias compounds with a curvature term, because the relevant marginal derivative is evaluated at a convex combination of the observed treatment and its conditional mean rather than at the observed treatment value itself.

These results situate residual-on-residual regression within the broader family of g-estimators of structural mean models, of which the linear no-effect-modification structural nested mean model is the simplest member.\cite{VansteelandtSjolander2016} Robinson's estimator coincides asymptotically with the two-step g-estimator that (i) fits the propensity score $\pi(W)$ by regressing the exposure on $W$, and (ii) regresses the outcome on the centered exposure $A - \pi(W)$ with adjustment for $W$, yielding an estimator that is doubly robust against misspecification of either the propensity or the conditional outcome regression.\cite{VansteelandtSjolander2016} AIPW and TMLE, which target the average treatment effect directly through the efficient influence function, do not share these heterogeneity biases.

\end{document}